\documentclass[structabstract]{aa}

\usepackage{graphicx}
\usepackage{txfonts}
\usepackage{natbib}
\bibpunct{(}{)}{;}{a}{}{,} 
\usepackage{amssymb}

\newcommand{\jastp}{J. Atmospheric \& Solar-Terrestrial Phys.}

\newcommand{\funits}{~$\rm erg~cm^{-2}~s^{-1}$}

\newcommand{\lengthunits}{~$R_\odot$}
\newcommand{\velunits}{~$\rm km~s^{-1}$}
\newcommand{\nofluxunits}{~$\rm cm^{2}~s^{-1}$}

\def\vec#1{\ensuremath{\mathbf{#1}}}

\newcommand{\qwav}{Q_{\mathrm{wav}}}

\usepackage{color}

\begin{document}
\title{Solar winds along curved magnetic field lines}
\author{B. Li\inst{\ref{inst1},\ref{inst2}} \and L.~D. Xia\inst{\ref{inst1}} \and Y. Chen\inst{\ref{inst1}}}
\institute{Shandong Provincial Key Laboratory of Optical Astronomy \& Solar-Terrestrial Environment,
  School of Space Science and Physics, Shandong University at Weihai, Weihai 264209, China
\email{bbl@sdu.edu.cn}
\label{inst1}
\and
State Key Laboratory of Space Weather, Chinese Academy of Sciences, Beijing 100190, China\label{inst2}}

\titlerunning{Solar winds along curved magnetic field lines}
\authorrunning{Li et al.}

\date{Received / Accepted}

\abstract
{
Both remote-sensing {measurements using the interplanetary scintillation (IPS) technique} 
     and in situ measurements {by the Ulysses spacecraft} show a bimodal structure
     for the solar wind at solar minimum conditions.
At present what makes the fast wind fast and the slow wind slow still remains to be answered.
While a robust empirical correlation exists between the coronal expansion rate $f_c$ of the flow tubes
    and the speeds $v$ measured in situ,
    further data analysis suggests that $v$ depends on more than just $f_c$.
}
{
We examine whether the non-radial shape of field lines, which naturally accompanies any non-radial expansion,
    could be an additional geometrical factor.
}
{We solved the transport equations incorporating the heating due to turbulent Alfv\'en waves
    for an electron-proton solar wind along curved field lines given by
    {{an analytical magnetic field}} model, representative of a solar minimum corona.
}
{The field line shape is found to influence substantially the solar wind parameters,
    reducing the asymptotic speed by up to $\sim 130$\velunits\, or by $\sim 28\%$ in relative terms,
    compared with the case neglecting the field line curvature.
This effect was interpreted in the general framework of energy addition in
    the solar wind:
    Relative to the straight case, the field line curvature
    enhances the effective energy deposition to the subsonic flow, resulting in
    a higher proton flux and a lower terminal proton speed.
}
{
Our computations suggest that the field line curvature could be a geometrical factor which, in addition to
    the tube expansion, substantially influences the solar wind speed.
Furthermore, at solar minima although the field line curvature unlikely affects the polar fast solar wind,
    it does help make the wind at low latitudes slow, thereby helping better reproduce the Ulysses measurements.
}
\maketitle

\section{INTRODUCTION}

Around solar minima the global distribution of the solar wind
    exhibits a simple bimodal structure:
    a band around the helioequator with an angular width $\sim 50^\circ$ of the slow wind (speed $v\lesssim 500$\velunits) is sandwiched between
    the high-latitude fast
    winds ($v\gtrsim 650$\velunits)~\citep[e.g., Plate 1 in][]{2000JGR...10510419M}.
This picture was established indirectly by interplanetary scintillation (IPS) measurements
    in 1974 and 1985, corresponding to the minimum of
    solar cycles (SC) 20 and 21, respectively~\citep{1990SSRv...53..173K}.
However, it became definitive only when Ulysses finished
    the first polar orbit which
    encompasses the minimum of SC 22~\citep{2000JGR...10510419M}.
The bimodal picture continued for the SC 23 minimum as shown by Ulysses during its
    third polar orbit~\citep{2008GeoRL..3518103M, 2009JGRA..11401109E}.

What makes the fast solar wind fast and slow wind slow?
In the context of Ulysses measurements, what produces the pronounced
    latitudinal variation of $v$?
In light of the well-established empirical relation between
    $v$ and the {coronal} expansion rate $f_c$
    of flow tubes~\citep{1990ApJ...355..726W},
    one may expect that the speed profile results from the differences in $f_c$ of tubes
    reaching different latitudes in interplanetary space.
Models that incorporate heating rates closely related to the magnetic field $B$ indeed suggest
    that varying the spatial distribution of $B$, or equivalently that of the expansion rate,
    can result in considerable changes in wind parameters
    (for a summary of geometrical models, see~\citeauthor{2005ESASP.592..159C}~\citeyear{2005ESASP.592..159C};
     for more recent models adopting empirical heating terms, see~\citeauthor{2011SSRv..tmp....3W}~\citeyear{2011SSRv..tmp....3W}).
For instance, the most recent models based on the anisotropic turbulence, which accommodate
     self-consistently the heating due to interactions of counter-propagating Alfv\'en waves,
     indicate that by varying the $B$ profile alone, the modeled latitudinal dependence of wind speed at 1~AU
     can agree rather well with the Ulysses
      measurements~\citep[Fig.12 in][]{2007ApJS..171..520C}.

While the $f_c - v$ relationship is strikingly robust, the significant spread in $v$ in a given $f_c$ interval
     means that $v$ is a function of more than $f_c$,
     and the angular distance $\theta_b$ of the field line footpoint to its nearest coronal hole boundary
     was proved a good candidate of this additional factor~\citep{2004JASTP..66.1295A}.
For a given $f_c$, the smaller $\theta_b$ is, the lower the speed.
As one naturally expects that
      the open field line becomes increasingly curved while its footpoint becomes closer to the coronal hole boundary,
      a natural alternative to $\theta_b$ would be the shape of field lines.
Surprisingly, the possible effect of field line curvature has not been examined.
Instead, the available turbulent-heating-based models tend to assume a radially directed flow tube,
      even though the non-radial tube expansion has been examined in considerable detail
      ~\citep[e.g.,][]{2002Ap&SS.282..447C, 2007ApJS..171..520C, 2011SSRv..tmp....3W}.

The aim of this study is to examine how
      the field line shape affects the solar wind parameters in general,
      and whether it helps reproduce the Ulysses measurements in particular.
It should be noted that while fully multi-dimensional models incorporating turbulent or empirical heating rates
      \citep[e.g.,][]{2000JGR...10512675U,2001SoPh..199..371C, 2004JGRA..10907103L,2007ApJ...654L.163C,  2009ApJ...690..902L, 2010ApJ...723..300F, 2010ApJ...725.1373V}
      certainly have taken account of both a non-radial expansion and the field line curvature,
      the two effects cannot be separated.
Therefore we choose to solve the transport equations along field lines
      in a prescribed magnetic field configuration typical
      of the solar minimum corona, thereby allowing us to switch on and off the effect of field line shape.
The solar wind model we employ incorporates a physically motivated boundary condition, distinguishes between electrons and protons,
      adopts reasonably complete energy equations including radiative losses, electron heat conduction, and the solar wind heating
      from parallel propagating turbulent Alfv\'en waves.

The paper is organized as follows.
In section~\ref{sec_model_des}, we give a brief overview of the model and
      describe how it is implemented.
The numerical solutions are then presented
      in section~\ref{sec_num_res}.
Finally, section~\ref{sec_conc} summarizes the results, ending with some
      concluding remarks.

\section{Model Description and Solution Method}
\label{sec_model_des}

We consider a time-independent electron-proton
    solar wind flowing in the meridional plane and
    incorporating turbulent heating from Alfv\'en waves.
Since the equations are quite well-known, we simply note that
    the mass or energy exchange between
    neighboring flow tubes is prohibited,
    permitting the decomposition of
    the vector equations \citep[e.g., Eqs.(2) to (7) in][and references therein]{2005JGRA..11012112L}
    into a force balance condition transverse to
    $\vec{B}$ and a set of conservation equations along it.
For mathematical details, please see the Appendix of~\citet{2006A&A...456..359L}.
In this study we replace the former with a prescribed
    magnetic field, and treat in detail only the conservation equations.
The Alfv\'en waves are assumed to propagate
    along $\vec{B}$, and to be dissipated at the Kolmogorov rate
    $\qwav = \rho \delta v^3/L_c$ where $\rho = n m_p$
    is the mass density with $n$ and $m_p$ being the number density and proton mass, respectively.
Moreover, $\delta v$ is the rms value of the wave velocity fluctuations,
    and $L_c$ the correlation length.
By standard treatment, $\qwav$ is entirely deposited to protons,
    and $L_c = L_{c0} (B_0/B)^{1/2}$ where
    the subscript $0$ denotes values at the field line footpoint at the Sun.
This form of dissipation and its variants, first proposed by~\citet{1986JGR....91.4111H},
    have been a common practice in solar wind modeling ever since~\citep[see e.g., the review by][]{2002JGRA..107.1147H}.

The conservation equations along an arbitrary magnetic line of force read
\begin{eqnarray}
&& \frac{\left(n v A\right)'}{A} =0,
      \label{eq_n}\\
&& v v' +\frac{p_p'}{\rho}
       +\frac{p_e'}{\rho}-\left(\frac{G M_\odot}{r}\right)'
       = \frac{F}{\rho}, \label{eq_v} \\
&& T_s'+(\gamma-1)T_s[\ln(vA)]' \nonumber \\
&& =\frac{\gamma-1}{nk_B v}\left[\frac{\delta E_s}{\delta t}+Q_s-L-\frac{\left(q_s A\right)'}{A}\right],
\label{eq_Ts}  \\
&&   \frac{\left(F_w A\right)'}{A}  + v F = -\qwav,
\label{eq_pw}
\end{eqnarray}
    where $v$ is the wind speed, $A\propto 1/B$ is the cross-sectional area of the flux tube,
    and the prime $'$ denotes the derivative with respect to the arc length $l$,
    measured from the footpoint of
    the field line.
The species pressures $p_s$ ($s=e, p$) are related to species temperatures $T_s$ via
    $p_s = n k_B T_s$ in which $k_B$ is the Boltzmann constant.
The gravitational constant is denoted by $G$,
    and $M_\odot$ the mass of the Sun.
Note that the gravitational acceleration is put in the form of a directional derivative.
In addition, the force density acting on the fluid is due to Alfv\'en waves, $F=-p_w'$  where
    $p_w=\rho \delta v^2/2$ is the wave pressure.
The adiabatic index $\gamma=5/3$,
    and $\delta E_s/\delta t$ represents the
    energy exchange rates of species $s$ due to Coulomb collisions with the other species.
To evaluate them, we take the Coulomb logarithm to be $23$.
For simplicity, the proton heat flux is neglected ($q_p = 0$), whereas
    the electron one is taken to be the Spitzer law
    $q_e = -\kappa T_e^{5/2}T_e'$
    where $\kappa = 7.8\times 10^{-7}$ {erg}~{K}$^{-7/2}$~{cm}$^{-1}$~s$^{-1}$.
Moreover, the heating rates applied to species $s$ is denoted by $Q_s$.
For the protons, one has $Q_p = \qwav$.
On the other hand, optically thin radiative losses are accounted for
    by the term $L$ which is included in the electron
    version of Eq.(\ref{eq_Ts}),
    and described by the parametrization scheme due to~\citet{1978ApJ...220..643R}.
Some electron heating is also included, expressed by
    $Q_e = \bar{Q}_{e} \exp(-l/l_d)$ where $\bar{Q}_{e}$ is the magnitude and $l_d$ the scale length of this heating.
For simplicity, here we assume that this electron heating is associated with no momentum deposition.
Equation~(\ref{eq_pw}) describes the wave evolution, in which $F_w = {p_w} \left(3v + 2v_A\right)$ is the wave energy flux density
    with $v_A = B/\sqrt{4\pi \rho}$ the Alfv\'en speed.

Given the governing equations, one may say a few words on the effect of
    the magnetic field line shape before actually solving them.
Consider the case, referred to as Case S, where we
    put $dl=dr$ in $'$ (note that $'$ is also involved in $q_e$)
    and $l=r-R_\odot$ in $Q_e$.
Here $r$ is the heliocentric distance and $R_\odot$ the solar radius.
As such, Case S differs from the original case, called Case C, only in that
    the field lines are straight in the former, but allowed to be curved in the latter.
It then follows that if the wind energetics is not self-consistently considered, say,
    if one prescribes an $r$-distribution of the temperatures $T_e$ and $T_p$,
    then Eqs.(\ref{eq_n}) and (\ref{eq_v}) indicate that no matter how curved
    the field line may be, the same $r$-profiles for $n$ and $v$ result.
This applies even if the waves are considered as long as the wave dissipation
    is neglected.
To understand this, one may readily derive from Eq.(\ref{eq_pw}) a conservation law which reads~\citep[e.g.][]{1997JGR...10214661H}
\begin{eqnarray}
 \frac{\left(\omega S A\right)'}{A} = -(1+M_A)\qwav,
\label{eq_wS_actflx}
\end{eqnarray}
     where $\omega {S} = 2 p_w {v}_A (1+M_A)$
     is the wave action flux density, and $M_A=v/v_A$ is the Alfv\'enic Mach number.
If $\qwav = 0$ then one finds that $\omega S A$ is a constant, enabling the wave pressure $p_w$ to be
     expressed as a functional of flow parameters as well as $B$.
As such multiplying Eqs.(\ref{eq_n}) and (\ref{eq_v}) by a factor
    $dl/dr$ does not invalidate the equal signs.
This means any two field lines that have identical radial distributions of $B, T_e$ and $T_p$ will yield
    identical radial distributions of $n, v$ and $p_w$, provided that the base values for $n$ and $\delta v$
    are the same for the two lines.
However, if the energy equation is explicitly considered, as we will do, and the adopted heating rates play
    a role in determining the solar wind mass flux, then the field line shape will likely play a role
    (see also the discussion section).

For the meridional magnetic field, we adopt the \citet{1998A&A...337..940B} model for which
     we use the reference values $[Q, K, a_1] =[1.5, 1, 1.538]$.
This set of parameters is supported by images obtained with the instruments
     on board {the SOlar and Heliospheric Observatory (SOHO) spacecraft} at solar minimum \citep[see e.g., Fig.1 in][]{1999SSRv...89....7F}, and
     has found extensive applications in solar wind models~\citep[e.g.,][]{2000A&A...360.1139H,2007ApJS..171..520C,2008ApJ...682..667L}.
Another parameter $M$ which controls the magnetic field strength is chosen
     such that $B$ at the Earth orbit is 3.5$\gamma$,
     consistent with Ulysses measurements~(\citeauthor{1995GeoRL..22.3317S}~\citeyear{1995GeoRL..22.3317S};
     also~\citeauthor{2011JASTP..73..277S}~\citeyear{2011JASTP..73..277S}).

\begin{figure*}
\centering
\includegraphics[width=0.95\textwidth]{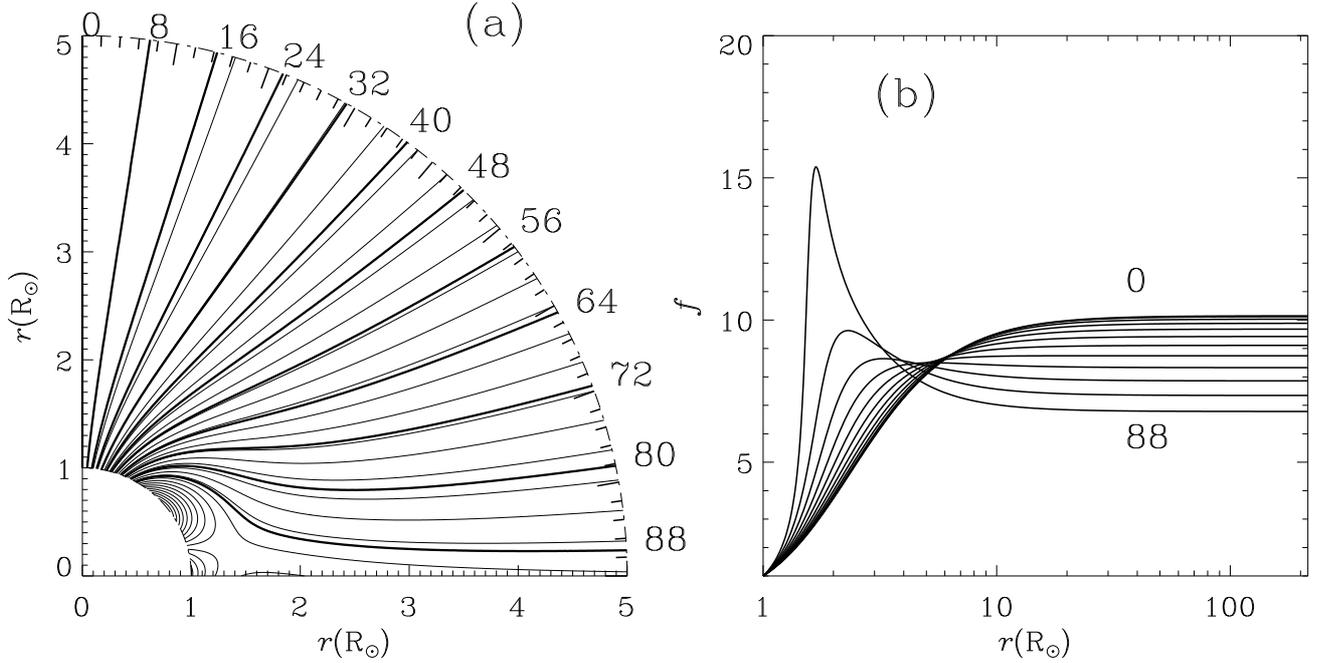}
\vskip 0.5cm
\caption{
Adopted meridional magnetic field configuration in the corona.
(a) A quadrant is shown where the magnetic axis points upward, and the thick contours
     delineate the lines of force for which the expansion factors $f=(A/A_0)(R_\odot/r)^2$ are plotted in (b) as a
     function of heliocentric distance $r$.
The field lines are labeled with the colatitude at which they intersect the Earth orbit (1~AU).
Here $A$ denotes the cross-sectional area of the flow tube, the subscript $0$ represents values at
     the inner boundary $R_\odot$.
}
\label{fig_mf}
\end{figure*}

The meridional magnetic field configuration is depicted in Fig.\ref{fig_mf}a,
     where the thick contours
     indicate the lines of force along which
     we compute the expansion factor $f= (A/A_0)(R_\odot/r)^2$, given in Fig.\ref{fig_mf}b.
The field lines are labeled with the colatitude at which the lines are located at 1~AU.
From Fig.\ref{fig_mf}a one can see that the open field lines occupy only a minor fraction of the solar surface,
     with colatitudes $\theta\le 30^\circ$ to be specific.
If identifying this area as the coronal hole (CH)~\citep[see e.g.,][]{2011SSRv..tmp....3W}, one can see that indeed when their footpoints
     approach the CH boundary, the open field lines become progressively more curved.
Furthermore, Fig.\ref{fig_mf}b shows that while in the inner corona, say at $r=1.7$\lengthunits, the expansion factor $f$ increases when the field lines
     become closer to the CH edge, the opposite is true at large distances, say {$r\gtrsim20$\lengthunits}.
Actually as was first pointed out by~\citet{1990ApJ...355..726W}, this tendency is due to the inclusion of the current sheet
     component in the magnetic field, and the low densities in the fast wind in interplanetary space
      were attributed to the the fact that
     flow tubes originating near the CH center underwent the most significant {net} expansion.

Given a $B$ profile along a field line, the governing equations
    are put in a time-dependent form,
    discretized onto a non-uniformly spaced grid,
    and then advanced in time from an arbitrary initial state by a fully implicit scheme~\citep{1997JGR...10214661H} until
    a steady state is reached, which is then taken as our solution.
For the solutions found, the mass, momentum, and energy are found to be conserved to better than $1\%$.
At the top boundary (1~AU), all the flow parameters are linearly extrapolated for simplicity.
At the base (1~$R_\odot$) the flow speed
     $v$ is determined by mass conservation,
     whereas the species temperatures are fixed at
     $T_{e0}=T_{p0}=T_0=5\times 10^5$~K, corresponding to mid-Transition Region.
Because the processes taking place below this level should influence the mass
     supply to the solar wind~\citep[e.g.,][]{1995JGR...10021577H,2002ApJ...566..562L},
     they should be taken into account
     and can be well represented by
     the Radiation Equilibrium Boundary (REB) condition~\citep{1988ApJ...325..442W, 2001ApJ...546..542L}.
The REB condition translates into $n_0 = 4.8\times 10^3 q_{e0}$ in cgs units
     for our chosen $T_0$.
As such the base density $n_0$ is not fixed but allowed to adjust to the downward
     heat flux, which in turn depends
     on how the fluid is heated close to the base.
Without the basal electron heating, the $T_e$ gradient and hence the electron heat flux density $q_{e0}$
     at the base are too small,
     leading to $n_0 T_0 \ll 1.6\times 10^{14}$~cm$^{-3}$ K, the value inferred from 
     {measurements with the Solar Ultraviolet Measurements of Emitted Radiation (SUMER) on SOHO}
     for coronal holes~\citep{1999JGR...104.9781W}.
The values adopted for $Q_e$ are $\bar{Q}_{e}=1.7\times 10^{-5}$~{erg~cm}$^{-3}$~s$^{-1}$, and $l_d=0.06$\lengthunits.
With such a short scale length, $Q_e$ rapidly gives way to the more gradual proton heating $\qwav$ for which
     we use a base wave amplitude $\delta v_0 = 27$\velunits\ and
     a base value for the correlation length $L_{c0}$ of $1.5\times 10^4$~km.
(Note that even though this $L_{c0}$ translates into approximately 0.02\lengthunits, it does not mean that the wave
     dissipation is extremely concentrated at the base.
A length scale comparable to $l_d$ involved in $Q_e$ can be given by $|\qwav/\qwav'|$, for which one finds a value
     ranging from 0.086\lengthunits\ at the base to 0.59\lengthunits\ at 2\lengthunits\ for the solution labelled case C
     in Fig.\ref{fig_ref_model}).

\section{NUMERICAL RESULTS}
\label{sec_num_res}

To examine the extent
     to which the wind parameters are affected by the field line shape,
     we will compare solutions incorporating the curvature effect (Case C) and
     solutions neglecting it (Case S).

\begin{figure}
\centering
\includegraphics[width=0.95\columnwidth]{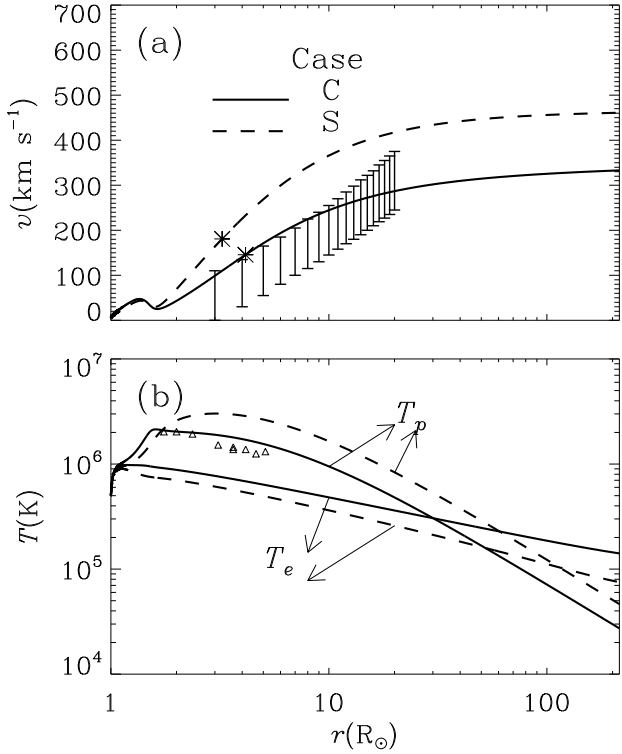}
\vskip 0.5cm
\caption{
Alfv\'en wave-turbulence-based solar wind models incorporating effects of field line shape.
Radial profiles are shown for
     (a) the solar wind speed $v$,
     and
     (b) the electron and proton temperatures $T_e$ and $T_p$.
The curved (straight) case in which the field line shape is considered (neglected) is given by
     the solid (dashed) lines.
In (a),
     the asterisks indicate the sonic point where $v$ equals the effective sonic speed $c_S$ defined by Eq.(\ref{eq_cs}),
     while the vertical bars represent wind speeds measured by tracking small density inhomogeneities
     (blobs) in SOHO/LASCO images
     as given by~\citet{2000JGR...10525133W}.
The triangles in (b) give the Hydrogen kinetic temperatures derived from H I Ly$\alpha$ measurements with SOHO/UVCS
    for a typical solar minimum streamer~\citep{2002ApJ...571.1008S}.
}
\label{fig_ref_model}
\end{figure}

Figure~\ref{fig_ref_model} presents the radial profiles of (a) the wind speed $v$ and
     (b) the electron and proton temperatures $T_e$ and $T_p$ for the solutions computed
     along the tube labeled $88$ in Fig.\ref{fig_mf}a.
For such a field line the effect of the field line curvature is almost maximal.
The solid and dashed curves correspond to Cases C and S, respectively.
The asterisks in Fig.\ref{fig_ref_model}a refer to the sonic point where $v$ equals the effective
     sound speed $c_S$ given by~\citep[e.g., Eq.(36) in][]{2000A&A...360.1139H}
\begin{eqnarray}
c_S^2 = \frac{k_B}{m_p}\left(T_e+T_p\right)+\frac{p_w}{\rho}\frac{1+3M_A}{2(1+M_A)} .
\label{eq_cs}
\end{eqnarray}
Moreover, in Fig.\ref{fig_ref_model}a the vertical bars give the range of wind speeds derived by
     tracking a collection of small inhomogeneities (the blobs) in
     {images obtained with the Large Angle Spectrometric COronagraph (LASCO) on board SOHO}~\citep{2000JGR...10525133W}.
The triangles in Fig.\ref{fig_ref_model}b
     represent the Hydrogen kinetic temperatures
     derived from the H I Ly$\alpha$ profiles {measured with the UltraViolet
Coronagraph Spectrometer (UVCS) on SOHO}
     for a typical solar minimum streamer~\citep{2002ApJ...571.1008S}.

The solution for Case C (solid curves) is in good agreement with available measurements.
For instance, the model yields at 1~AU the following parameters:
     $v_{E}=333$\velunits, $(nv)_E=3.65\times 10^8$\nofluxunits,
     $T_e=1.4\times 10^5$~K and $T_p=2.7\times 10^4$~K,
     all of which match the in situ measurements
     by both Helios~\citep{1990pihl.book...99S} and
      Ulysses~\citep{2000JGR...10510419M,2008GeoRL..3518103M,2009JGRA..11401109E}
     of the low latitude slow winds.
Here the subscript $E$ denotes the Earth orbit.
Figure~\ref{fig_ref_model}a indicates that the wind speed starts with $6.4$\velunits\ at the base,
     in line with the Doppler speed measured with Ne VIII~\citep[e.g.,][]{2003A&A...399L...5X} whose formation temperature
     is close to our base temperature.
The speed profile is not monotonic but possesses a local minimum at $1.63$\lengthunits\ where $v=25.1$\velunits.
The sonic point, denoted by $r_C$, is located at $4.15$\lengthunits.
Between $3$ and $20$\lengthunits\ the computed speeds are in line with the blob measurements.
Moreover, for $r \lesssim 3$\lengthunits\ the modeled $v$ is no higher than $98.8$\velunits\, which would be below
     the sensitivity of the Doppler dimming of H I Ly$\alpha$~\citep{2006A&ARv..13...31K}, and in this sense
     compatible with the fact that there is not much empirical knowledge about the proton speed
     in the inner portion of the slow wind.
Moving to Fig.\ref{fig_ref_model}b one may see that due to the applied electron heating both
     $T_e$ and $T_p$ increase near the base.
At $1.19$\lengthunits\ the electron temperature $T_e$ attains its maximum of $9.8\times 10^5$~K.
Below this height, $T_e$ and $T_p$ stay close to each other due to Coulomb coupling.
However, above this height while $T_e$ decreases monotonically with distance, $T_p$
     further increases to $2.14\times 10^6$~K at $1.6$\lengthunits\ beyond which
     $T_p$ declines with a slope steeper than $T_e$ and is overtaken by $T_e$ at $30.1$\lengthunits.
In the range from $1.75$ to $5.11$\lengthunits\ the modeled proton temperature reproduces rather satisfactorily
     the H I measurements, indicating the heating mechanism we employ performs well in
     heating the inner solar wind.

Compared with Case C, Case S yields at $1$~AU a significantly higher proton speed and a substantially lower proton flux density.
The parameters at 1~AU now read
    $v_E = 461$\velunits, $(nv)_E = 3.14\times 10^8$\nofluxunits,
    $T_e = 7.5\times 10^4$~K,  and $T_p=4.65\times 10^4$~K.
In addition, Case S yields a sonic point at $r_C=3.24$\lengthunits, closer to the Sun than in Case C.
The differences between the temperatures in the two cases can be readily understood by examining
    Eq.(\ref{eq_Ts}).
A higher speed in Case S leads to a more prominent adiabatic cooling (the second term on the left-hand-side)
    which accounts for a steeper, negative $T_e$ gradient and hence a lower asymptotic $T_e$;
    on the other hand, while the same also happens to protons, the reduced proton flux results in
    a higher $Q_\mathrm{wav}/(nv)$
    which leads to a steeper, positive $T_p$ gradient and eventually a higher asymptotic $T_p$.
However, the question remains as to how neglecting the curvature of the field line leads to the changes in
    the proton speed and flux?
Actually this is understandable if one follows the comprehensive discussion by \citet{1980JGR....85.4681L}:
    the more energy deposited in the subsonic region of the flow, the higher the proton flux;
    and the larger energy deposition per particle, the higher the terminal speed.
Let us first examine the subsonic physics,
    from $R_\odot$ to the sonic point $r_C$.
The contribution from the electron heating, measured as an energy flux density scaled to
    1AU $\int_{R_\odot}^{r_C} Q_e(A/A_E)dl$,
    increases only slightly from 0.3 in Case C to 0.337 in Case S
    (here and hereafter the energy flux densities are in \funits).
However, the effect of energy addition to protons, which eventually derives from
    Alfv\'en waves since
    $\int_{R_\odot}^{r_C} (Q_p+v F)(A/A_E)dl = F_w A/A_E|_{R_\odot}^{r_C}$, reduces more significantly
    from 1.089 in Case C to 0.931 in Case S.
And this reduction is a direct consequence of the field line shape.
To see this, we note that the wave action equation~(\ref{eq_wS_actflx}) is
    a proper starting point, since its right-hand-side (RHS) evaluates
    the consequence of the ``genuine'' dissipation while its LHS involves
    the wave ponderomotive force which also contributes to the wave energy loss.
Without dissipation, we have seen that the field line shape is unable to make the {\it radial}
    distributions different in Case~C from those in Case~S, even when the ponderomotive force is allowed for.
Expressing Eq.(\ref{eq_wS_actflx}) in the integral form yields
\begin{eqnarray}
\left.\left(\omega S \frac{A}{A_E}\right)\right|_{R_\odot}^{r_C}  = - \int_{R_\odot}^{r_C}(1+M_A)\frac{A}{A_E} Q_{\mathrm{wav}} dl.
\label{eq_ws_int}
\end{eqnarray}
Now that in general $dl/dr>1$, which is particularly true in the inner flow where $M_A \ll 1$,
     the RHS of Eq.(\ref{eq_ws_int}) means that $\qwav$ of similar amount
     effectively contributes more to the wave energy reduction in the curved
     than in the straight case.
Now let us discuss the asymptotic speeds.
The net energy added to the solar wind,
     i.e., the electron heating plus wave dissipation
     plus enthalpy flux subtracted by
     radiative losses and the net electron heat flux flowing out of the computational domain,
     yields an energy flux scaled to 1~AU of 1.495 (1.553) in Case C(S).
Divided by the proton flux density, this gives
     0.41 (0.495)$\times 10^{-8}$~{erg} per particle in Case C(S),
     thereby explaining the higher terminal speed in Case S.

Now return to the question that, besides the tube expansion, what geometrical factor(s) may also affect
     the solar wind speed $v$.
The comparison between the straight and curved cases shows that the field line shape could be one.
In this regard, the physical basis for the empirical relation found by \citet{2004JASTP..66.1295A}
     may be simply that $\theta_b$, which measures how far the open field line footpoint
     is away from the coronal hole boundary, also characterizes the field line shape.

\begin{figure}
\centering
\includegraphics[width=0.95\columnwidth]{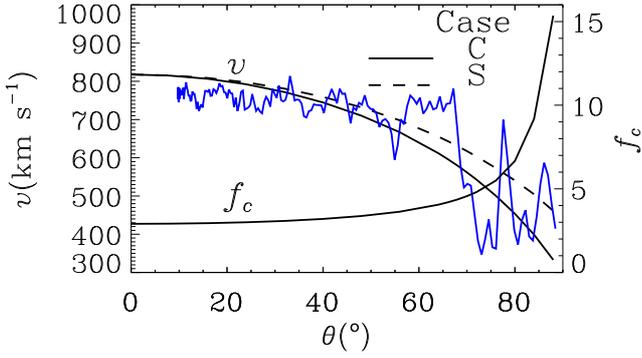}
\vskip 0.5cm
\caption{
Distribution with colatitude $\theta$ of the computed solar wind speed $v$ at 1~AU and
    the {coronal} expansion factor $f_c$ of the flow tube intersecting the Earth orbit at $\theta$.
The model result incorporating (neglecting) the field line shape, called Case C (S), is given by the solid (dashed) line.
They are obtained by varying the flow tube alone but keeping all other parameters unchanged.
Cases C and S both take the radial variation of magnetic field strength into account, and differ only in whether
    the field line shape is considered.
The blue continuous curve gives the daily averaged speed measurements by Ulysses during the first half
of its latitude scan from Sep 12 1994 to Mar 4 1995.
As for $f_c$, it is taken to be the value of $f$ at a heliocentric distance of $1.7$\lengthunits.
}
\label{fig_lat_profile}
\end{figure}

What role does the shape of field lines play in shaping the latitudinal profile of the solar wind speed at 1~AU?
This is examined by Figure~\ref{fig_lat_profile} which compares the speed profile in Case C (the solid curve)
     and in Case S (the dashed curve).
The model results are obtained by varying only the flow tube along which
    the governing equations are solved,
    but keeping everything else untouched.
In addition to the speed $v$ at 1~AU, Fig.\ref{fig_lat_profile} also displays as a function
    of colatitude $\theta$ the {coronal} expansion factor $f_c$ of the flow tube reaching 1~AU at $\theta$.
Here $f_c$ is defined as $(A/A_0)(R_\odot/r_{\mathrm{cor}})^2$, where $r_{\mathrm{cor}}$
    is taken to be $1.7$\lengthunits\ (see Fig.\ref{fig_mf}b).
For comparison, the solid curve in blue gives the daily averages of Ulysses measurements
    of the wind speed during the first half
    of the fast latitude scan from Sep 12 1994 to Mar 4 1995.
From Fig.\ref{fig_lat_profile} one can see that while the speed $v$ in both models
    is inversely correlated with $f_c$, the results in Case C provide
    a stronger latitudinal variation.
As a matter of fact, the reduction in the wind speed due to
    the field line shape is larger than $10\%$ throughout the latitudinal range
    $\theta \gtrsim 70^\circ$, with
    the largest relative difference being $27.8$\% occurring close to the equator.
Furthermore, without further tweaking of the heating parameters, say varying the base values of
    wave amplitude $\delta v_0$ and correlation length $L_{c0}$ from tube to
     tube as was done by e.g.,~\citet{2004JGRA..10907103L},
    the modeled latitudinal dependence for $\theta\lesssim 70^\circ$  is too strong to reproduce
    the measured profile for the fast wind.
Despite this, the low latitude portion of Fig.\ref{fig_lat_profile} indicates that in addition
    to a more pronounced coronal expansion rate, the field line curvature is also an important
    factor that makes the slow wind there slow.

\section{Summary and Concluding Remarks}
\label{sec_conc}
Both remote-sensing {measurements using the interplanetary scintillation (IPS) technique}
    and in situ measurements {by the Ulysses spacecraft} show a latitudinal gradient in the wind properties,
    the speed in particular, at solar minimum conditions
    \citep{1990SSRv...53..173K, 2000JGR...10510419M, 2008GeoRL..3518103M}.
What makes the fast wind fast and the slow wind slow still seems elusive.
While there exists a robust empirical correlation between the coronal expansion rate $f_c$ of the flow tubes
    and the speeds $v$ measured in situ~\citep{1990ApJ...355..726W},
    further data analysis suggests that $v$ depends on more than just $f_c$~\citep{2004JASTP..66.1295A}.
We examine whether the non-radial shape of field lines, which naturally accompanies any non-radial expansion,
    could be an additional geometrical factor.
To this end, we solved the transport equations for an electron-proton solar wind, which incorporate the heating due to
    turbulent Alfv\'en waves dissipated at the Kolmogorov rate, along curved field lines given by
    the~\citet{1998A&A...337..940B} model, representative of a solar minimum corona.
The shape of field lines was found to substantially influence the solar wind parameters,
    reducing the terminal speed by as much as $\sim 130$\velunits\, or up to $28$\%, relative to the straight case where field line curvature is neglected.
And this effect was interpreted in the general framework of energy addition in
    the solar wind by~\citet{1980JGR....85.4681L}:
    Compared with the straight case, even though the wave dissipation rates may be similar,
    the field line curvature
    enhances the effective wave energy deposition in the subsonic region of the flow, resulting in
    a higher proton flux and a lower terminal proton speed.
Our results suggest that the experimental finding by~\citet{2004JASTP..66.1295A} may be interpreted in view of the fact
    that flow tubes having identical coronal expansion rates may differ substantially in their curvature.
On the other hand, at solar minima even though the field line curvature unlikely affects the polar fast solar wind,
    it does considerably help the wind at low latitudes become slow, thereby helping better reproduce the Ulysses measurements.

Could this effect have to do with the particular boundary conditions or the energy deposition mechanism employed here?
This proves very unlikely.
We have tested cases where the inner boundary is placed at the lower transition region where $T=10^5$~K, or
    the ``coronal base'' where $T=10^6$~K.
Adjusting the heating parameters such that realistic values can be obtained for the proton flux and asymptotic speed,
    and comparing solutions with field line curvature switched on and off, the conclusion
     is the same.
On the other hand, using a different heating function where $Q_p$ is given by the dissipation of some unspecified mechanical
    energy flux at a constant scale length, the conclusion is the same.
Moreover, this effect is not related to the heat conduction, even
    though the effective electron heat conductivity is reduced by a factor
    $dr/dl$ in the curved case.
Actually, turning off the electron heat flux, in which case we have to fix the
    base density though,
    we reach the same conclusion.
All of these additional computations corroborate our interpretation that the changes brought forth by the field line shape
    are simply related to the differential arc-length $dl$ involved in Eq.(\ref{eq_ws_int}), whereby the more curved the field lines are
    in the subsonic region, the more wave energy reduction there, hence the higher the proton flux, and
    with the total input energy flux barely changing, the lower the terminal proton speed.
The effect of field line shape therefore should happen in quite general situations where
    the wave dissipation rate $\qwav$ is not directly proportional to the spatial derivative of background parameters
    {\it and} the energy deposition in the subsonic region plays a significant role
    in determining the solar wind flux.
From this perspective one may expect that the effect of field line curvature should not be restricted to
    flow tubes rooted at any particular latitude on the Sun.
Instead, it should play a role for any flow tube that is curved.
{Moreover, this effect should also exhibit some influence on the flow if one employs the anisotropic
    turbulence treatment of the Alfv\'en waves to incorporate the finite-wavelength non-WKB effects,
    such as being actively pursued at present~\citep{2007ApJS..171..520C,2010ApJ...708L.116V}.
Compared with a radial one, a curved line of force is likely to enhance the non-WKB effect (i.e., reflection) in
    the near-Sun region~\citep[Fig.16 in][]{1980JGR....85.1311H}.
However, to predict what happens next is not as straightforward as was done here.
Neglect for the moment the wave dissipation, i.e., 
    the waves interact with the flow only via their ponderomotive force.
The enhanced reflection tends to lead to a reduction of this force, and hence to
    a reduced mass flux as well as a lower terminal speed.
Unlike the dissipationless case discussed in Section~\ref{sec_model_des}, this tendency exists
    both when the wind temperatures are prescribed~\citep[Fig.10 in][]{1994ApJ...430..387M}
    and when the wind energetics is self-consistently treated~\citep[Fig.4 in][]{2008ApJ...682..667L}. 
The case with turbulent dissipation is considerably more complicated.
The enhancement of wave reflection in the subsonic flow means an enhancement of the ingoing component of the waves.
With the turbulent dissipation trying to diminish both outgoing and reflected components to similar extents,
    a substantial fraction of the reflected component may end up in the supersonic portion of the flow,
    and its consequent dissipation may lead to a higher terminal speed with mass flux barely altered.  
However, this is not the whole story since the ponderomotive force is altered in a similar fashion to
    the dissipationless case.
As such the net outcome has to be told by a detailed numerical study, which is beyond the scope of this paper.
Nonetheless, it seems fair to say} while trying to reproduce the Ulysses measurements, especially those of the slow solar
    wind at low latitudes,
    the field line shape has to be accounted for.

That said, let us say a few words on the situations where the curvature
    effects are unlikely important.
This may happen, for instance,
1) in the spectral erosion scenario where the solar wind is heated by high frequency Alfv\'en waves (with frequencies up to 10~kHz) directly launched
    by chromospheric magnetic reconnections, since now the dissipation rate is proportional to the directional derivative
    of the proton gyro-frequency and therefore the background magnetic field strength~\citep{1997A&A...319L..17M,2000A&A...360.1139H}.
2) in the scenario for the solar wind generation where both the energy $F_{w0}$
    and mass injection $F_{m0}$ rates are fixed~\citep[e.g.,][]{2003JGRA..108.1157F, 2005Sci...308..519T}.
   To illustrate this, consider the scenario by \citet{2003JGRA..108.1157F}
       where the energy and mass injection occurs at
      some level where the temperature is of the order 1~MK, which means,
        roughly speaking, the contributions like enthalpy, heat flux and radiative losses may be neglected when
        considering the overall energy balance between the injection point and 1~AU.
   Evidently the mass flux density at 1~AU is simply $F_{m0} A_0/A_E$, and
       from energy conservation follows that the terminal proton
       speed is approximately $\sqrt{2F_{w0}/F_{m0}-G M_\odot/R_\odot}$.
   As such, the terminal mass flux density is determined by the {net} expansion of the flow tube from its footpoint
       to 1~AU, but the speed is related to neither the tube expansion nor the field line shape.

\begin{acknowledgements}
{We thank the referee (Dr. Andrea Verdini) for his comments which helped us clarify the possible effects of
    the magnetic field line shape on a turbulence-heated solar wind where the non-WKB effects are considered. }
The Ulysses data are obtained from the CDAWeb database.
The SWOOPS team (PI: D.~J.~McComas) is gratefully acknowledged.
This research is supported by the grant NNSFC 40904047,
     and also by the Specialized Research Fund for State Key Laboratories.
YC is supported by grants NNSFC 40825014 and 40890162,
   and LDX by NNSFC 40974105.
\end{acknowledgements}

\end{document}